\documentstyle[aps,epsf]{revtex}  
\begin{document}        
\newcommand {\ecm} {E$_{\,\mathrm{c.m.}}$}
\newcommand {\epem}     {e$^+$e$^-$}
\newcommand {\mnch} {$\langle n_{\,\mathrm{ch.}} \rangle$}
\newcommand {\gluglu} {gg}
\newcommand {\qbar} {$\overline{\mathrm{q}}$}
\newcommand {\qq} {q$\overline{\mathrm{q}}$}
\newcommand {\gincl} {g$_{\,\mathrm{incl.}}$}
\newcommand {\ejet} {$E_{\,\mathrm{jet}}$}
\newcommand {\xee} {$x_{E}$}
\newcommand {\nch} {$n_{\,\mathrm{ch.}}$}
\newcommand {\mysph} {$|y|$}
\newcommand {\rchysphone} {r_{\,\mathrm{ch.}}(\mmysph\leq 1)}
\newcommand {\rch} {r_{\,\mathrm{ch.}}}
\newcommand {\ysph} {$y$}
\newcommand {\mmysph} {|y|}
\newcommand {\cacf} {C$_{\mathrm{A}}$/C$_{\mathrm{F}}$}
\newcommand {\durham}   {$k_\perp$}

\baselineskip 14pt
\title{Recent QCD Results from LEP-1 and LEP-2}
\author{J. William Gary}
\address{Department of Physics, University of California, Riverside CA 92521}
\maketitle              

\begin{abstract}        
A summary is given of some recent QCD results from LEP.
For LEP-2,
the topics include event shape measurements,
determinations of $\alpha_S$,
and measurements of the charged particle
multiplicity distribution at the
recently completed run at {\ecm}=189~GeV.
For LEP-1,
the topics presented are
a test of the flavor independence of 
$\alpha_S$ and a study of gluon jets using a
hemisphere definition to correspond
to analytic calculations.
For the combined LEP data samples,
the topics include a
test of power law corrections for hadronization effects
and the running of~$\alpha_S$.
\
\end{abstract}   	

\section{Introduction}

In the LEP-1 period of data collection,
from 1989 to 1995,
about 170~pb$^{-1}$ of data were collected by each LEP
experiment at energies near the mass of the Z$^0$ boson,
yielding approximately $4\times 10^6$ hadronic annihilation
events per experiment.
So far at LEP-2,
which began in 1996 and is scheduled to run through 2000,
a total of 270~pb$^{-1}$ of data has been collected at
center-of-mass (c.m.) energies, {\ecm},
of 161, 172, 183 and 189~GeV.
Although the luminosity at LEP-2 is large,
the event rate is small:
the corresponding numbers of QCD events are only about
400, 240, 1300 and 3000 per experiment.
By ``QCD event,''
it is meant a hadronic annihilation event produced
through the s-channel decay of a virtual Z$^0/\gamma$
to a quark-antiquark pair,
in which there is minimal initial-state photon radiation
so that the hadronic system carries near to 
the full c.m. energy value.
Besides the LEP-1 and LEP-2 data,
LEP ran at an energy of about 133~GeV
in 1995 and 1997:
thus well above the Z$^0$ mass but below the
threshold for W$^+$W$^-$ production.
This data is sometimes referred to as LEP-1.5.
About 10~pb$^{-1}$ of data were collected at LEP-1.5,
yielding about 700 QCD events.
The LEP-2 data provide the possibility to study QCD
at the highest available {\epem} energies,
where uncertainties from hadronization and unknown
higher order terms in perturbative expressions for
experimental observables are relatively small.
The large statistics of the LEP-1 data 
allow tests and measurements not possible with
other data samples.
The LEP data together allow the energy evolution
of QCD quantities to be studied.
In the following,
we present a summary of some recent work in QCD
performed using the LEP data.
We first present preliminary results from the
recently completed run at 189~GeV.
Following this,
we discuss some unique and precise
tests of QCD made using LEP-1 data.
Last,
we present results using the combined LEP data samples
to test QCD predictions for the energy scaling of
several quantities.

\section{Results from 189~GeV}
\label{sec-189}

\begin{figure}[ht]
\vspace*{-1.2cm}
\begin{center}
\begin{tabular}{cc}
\epsfxsize=2.5 truein
\epsffile[225   150 755 750]{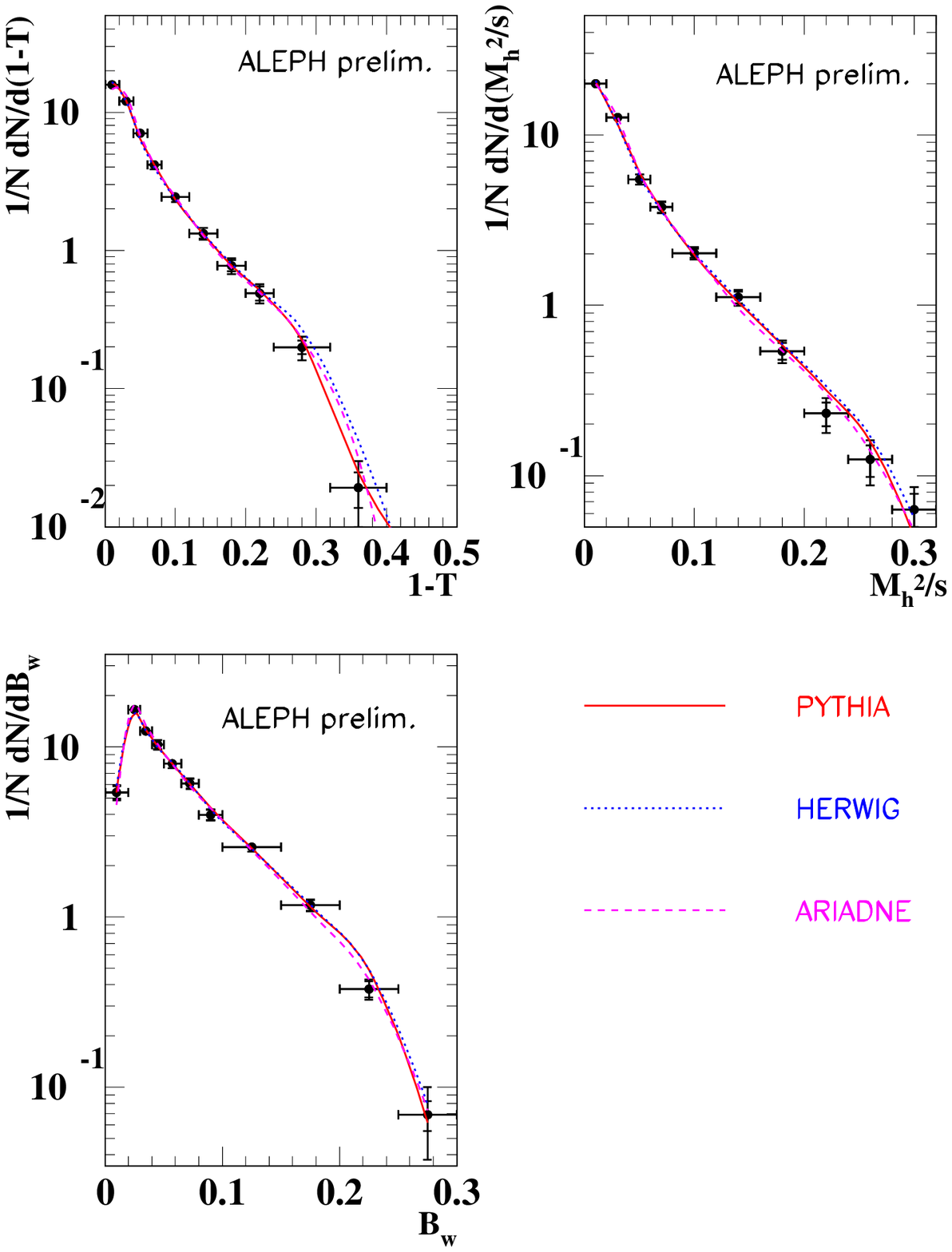} &
\epsfxsize=1.1 truein
\epsffile[225   180 455 280]{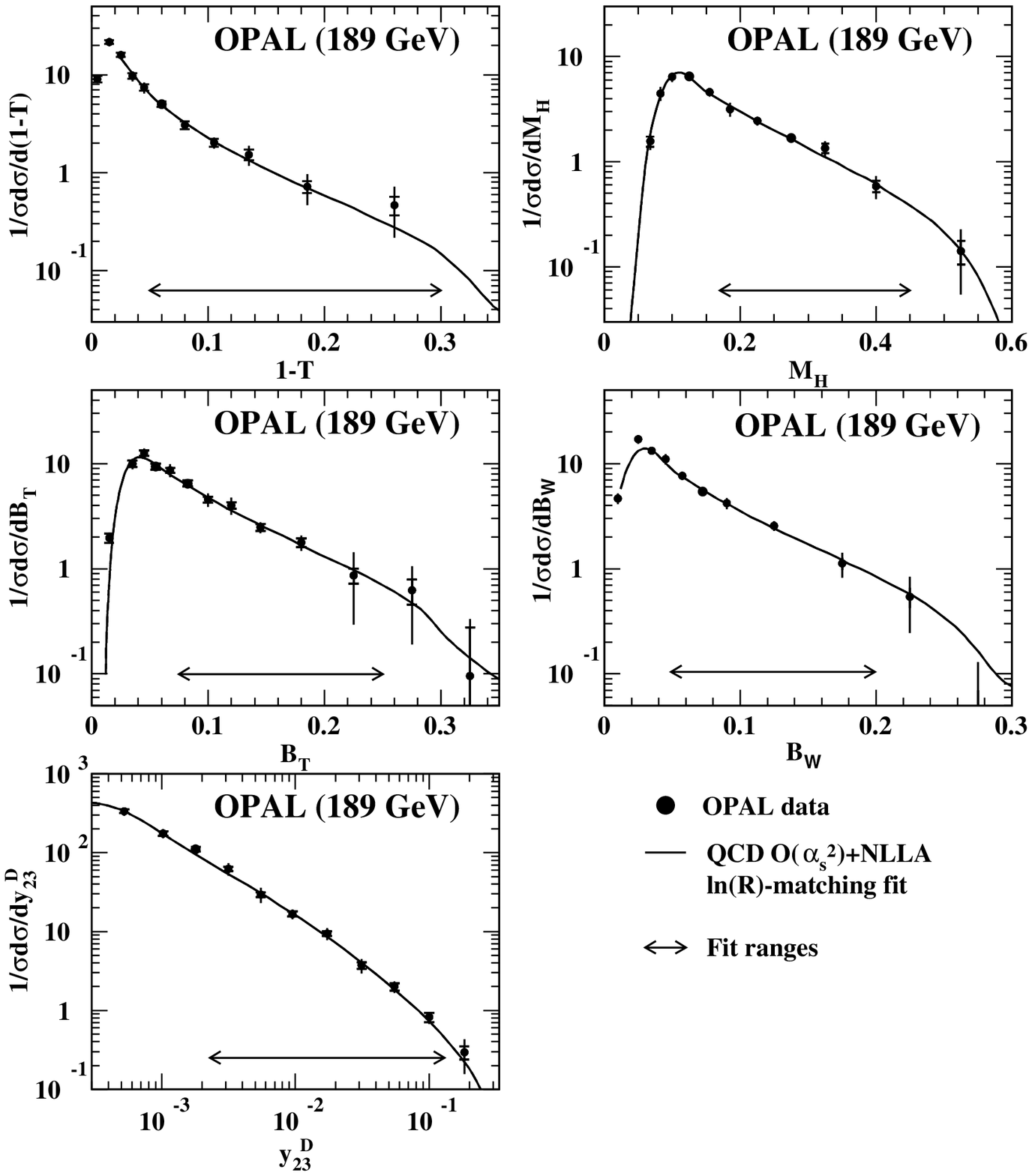} \\
\end{tabular}
\end{center}
\caption[]{
\label{fig-alephes}
\small Measurements of event shape variables at 189~GeV
from ALEPH (left)~\cite{bib-lancon98}
and OPAL (right)~\cite{bib-plane98}.
The ALEPH results are shown at the hadron level in comparison
to the predictions of QCD parton shower Monte Carlo programs.
The OPAL results are shown at the parton level;
the solid curves show the results of a fit of
${\cal{O}}(\alpha_S^2)$+NLLA calculations to the data.
}
\end{figure}

LEP ran at 189~GeV during 1998.
Each LEP experiment
collected a data sample of about 190~pb$^{-1}$,
more than the integrated luminosity collected during
the entire LEP-1 period.
The most basic QCD test which can be performed using
these data is to examine variables
which measure the distribution of particle energy and momenta
and to compare them to the predictions of QCD Monte Carlo
event generators.
Standard ``event shape'' variables used for this purpose
are \mbox{Thrust T}~\cite{bib-thrust}, 
Heavy Jet Mass M$_{\mathrm{H}}$~\cite{bib-mh},
Jet Broadening Variables B$_{\mathrm{W}}$
and B$_{\mathrm{T}}$~\cite{bib-btbw},
and y$_{23}^{\mathrm{D}}$ (sometimes referred to
as y$_{3}$ or D$_{2}$).
Thrust is defined as 
$$
\mathrm{T} = max\left(\frac{{\displaystyle \sum_{i=1,N}}
    \vec{p}_i\cdot\hat{n}_{\mathrm{T}}}
   {{\displaystyle \sum_{i=1,N}}\left|\vec{p}_i \right|}\right) \;\;\;\; ,
$$
where the thrust axis $\hat{\mathrm{n}}_{\mathrm{T}}$
is the unit vector which maximizes T, as indicted.
The sum is over the particles in the event,
with $\vec{\mathrm{p}}$ the particle momentum.
The quantities M$_{\mathrm{H}}$, B$_{\mathrm{W}}$
and B$_{\mathrm{T}}$ are defined by dividing events into
hemispheres using the plane perpendicular to 
$\hat{\mathrm{n}}_{\mathrm{T}}$:
M$_{\mathrm{H}}$ is the larger of the two 
hemisphere invariant mass values,
while B$_{\mathrm{W}}$ and B$_{\mathrm{T}}$
are defined by B$_{\mathrm{W}}$=max(B$_1$,B$_2$)
and B$_{\mathrm{T}}$=B$_1$+B$_2$ with
$$
   {\mathrm{B}}_{j} = \frac{{\displaystyle \sum_{i\in j}}
       \left|\vec{\mathrm{p}}_i\times
        \hat{\mathrm{n}}_{\mathrm{T}} \right|}
      {{\displaystyle\sum_{i=1,N}}\left|\vec{\mathrm{p}}_i\right|}
$$
with the index j=1,2 referring to the hemisphere and where the
sum in the numerator is over the particles in hemisphere~$j$.
Last, y$_{23}^{\mathrm{D}}$ is the resolution value at which an event
changes from being classified as a two-jet event to being
classified as a three-jet event using the {\durham}
(``Durham'') recombination jet algorithm~\cite{bib-durham}.
In Fig.~\ref{fig-alephes} (left),
ALEPH measurements of
1-T, (M$_{\mathrm{H}}$/{\ecm})$^2$ and B$_{\mathrm{W}}$
at 189~GeV
are shown in comparison to the predictions of the
Pythia~\cite{bib-pythia}, 
Herwig~\cite{bib-herwig} 
and Ariadne~\cite{bib-ariadne} Monte Carlo multihadronic
parton shower event generators.
The parameters of the event generators were tuned
using Z$^0$ data.
The event generators are seen to describe the 189~GeV
data well,
demonstrating that the energy evolution of the variables
is as expected from QCD.

The event shape variables defined above have an importance beyond
that of comparison to Monte Carlo predictions
in that they are stable under the
emission of soft and collinear radiation,
allowing them to be calculated perturbatively.
Calculations of these variables have been performed
to two loop order,
corresponding to ${\cal{O}}(\alpha_S^2)$
in {\epem} annihilations.
In addition to the ${\cal{O}}(\alpha_S^2)$ calculations,
the leading and next-to-leading logarithmically
divergent terms have been summed to all orders
into analytic functions,
an approximation known as~NLLA.
The combination of the two calculations yields predictions
valid to ${\cal{O}}(\alpha_S^2)$+NLLA,
the most complete theoretical description 
of event shape variables which is currently available.
In Fig.~\ref{fig-alephes} (right),
OPAL measurements of event shapes at 189~GeV are shown
after correction for the effects of hadronization
(the hadronization corrections are derived from
the Monte Carlo programs).
The solid curves in Fig.~\ref{fig-alephes} (right) show the results
of a fit of the ${\cal{O}}(\alpha_S^2)$+NLLA predictions
to the data.
The preliminary value of $\alpha_S$ extracted from this fit is
$0.106\pm0.001$(stat.)$\pm0.005$(syst.)~\cite{bib-plane98},
for which the largest systematic uncertainties
are from the hadronization corrections,
the matching of the ${\cal{O}}(\alpha_S^2)$ and NLLA calculations,
and the ambiguity in the value to choose for the renormalization scale.
L3 and ALEPH have also reported preliminary results
for $\alpha_S$ at 189~GeV,
using a similar technique.
The ALEPH and L3 results,
$\alpha_S$(189~GeV)=$
0.110\pm0.001$(stat.)$\pm0.003$(syst.)~\cite{bib-lancon98},
and
$0.1082\pm0.0028$(exp.)$\pm0.0052$(theo.)~\cite{bib-clare98},
can be combined with the OPAL value to yield
$\alpha_S$(189~GeV)=$0.108\pm0.001$(stat.)$\pm0.004$(syst.),
which is significantly smaller than the result
$\alpha_S$(91~GeV)=$0.121\pm0.003$(stat.+syst.)
obtained from the ratio of the hadronic to the leptonic
decay widths of the Z$^0$ measured at LEP-1~\cite{bib-alphaslep1}:
this is an example of the running of $\alpha_S$ between
LEP-1 and LEP-2.
The LEP combined result given above is obtained by taking the
statistical uncertainties as uncorrelated and the
simple mean of the systematic uncertainties.

Preliminary results for the
mean charged particle multiplicity
of QCD events at 189~GeV, {\mnch}(189~GeV),
have been presented by ALEPH and OPAL:
the results are
$27.37\pm0.20$(stat.)$\pm0.27$(syst.)~\cite{bib-lancon98}
and
$26.94\pm0.17$(stat.)$\pm0.41$(syst.)~\cite{bib-plane98},
respectively,
which can be combined to yield
$27.12\pm0.13$(stat.)$\pm0.34$(syst.).
This value is substantially larger than the value
measured at 91~GeV:
{\mnch}(91~GeV)=$21.00\pm0.20$(stat.+syst.).
A comparison of the 189~GeV result to Monte Carlo
predictions and a test of the energy scaling of {\mnch}
are presented below in section~\ref{sec-escaling}.

\section{Results from LEP-1}
\subsection{Flavor independence of {$\boldmath \alpha_S$}}

In QCD,
the strong interaction is flavor blind,
i.e. gluons couple with equal strength to
quarks of all flavors.
{\epem} colliders are well suited to test this aspect
of the Standard Model because flavor tagging techniques
and $\alpha_S$ measurements are both well developed areas.
The procedure is to tag the event flavor ``$f$'' in
{\epem}$\rightarrow$q$_f$$\overline{\mathrm{q}}_f\rightarrow hadrons$
events,
where the flavor tags are $f$=uds, c or b,
with uds an undifferentiated sample of light quark (uds) events.
In a recent OPAL study~\cite{bib-opalalphasf},
uds and b events are identified using the signed
impact parameter values of charged tracks with respect to
the primary interaction point, $b$,
since the distribution of this variable is strongly skewed
towards positive values for b events,
and to a lesser extent for c events,
but not for uds events.
By requiring that there be {\it no track} in an event
with $b/\sigma_b$$>$2.5,
where $\sigma_b$ is the uncertainty of $b$,
a uds sample purity of 86\% is obtained.
By requiring that there be {\it at least five tracks}
with $b/\sigma_b$$>$2.5,
a b sample with a purity of 96\% is selected.
c~events are identified by requiring the presence
of high energy D$^{*\pm}$ mesons,
yielding a sample purity of~55\%.
$\alpha_S$ is measured in the flavor tagged
samples using event shape variables such as are discussed
above in section~\ref{sec-189}.
The results for the flavor independence of $\alpha_S$
are presented in the form
$$
\frac{\alpha_S^b}{\alpha_S^{uds}} \;\;\; ; \;\;\;
\frac{\alpha_S^c}{\alpha_S^{uds}}\;\;\; ,
$$
where $\alpha_S^f$ is the strong coupling strength
measured for flavor sample~$f$.
The results are presented in this form
so that the main uncertainties,
from hadronization and the renormalization scale,
partially cancel in the ratios.
Thus the flavor independence of $\alpha_S$ can be
measured with greater precision than $\alpha_S$ itself.
\begin{figure}[t]
\vspace*{-1.7cm}
\begin{center}
\epsfxsize=2.6 truein
\epsffile[-625   50 55 650]{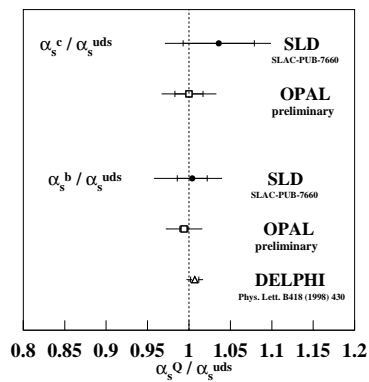}
\end{center}
\caption[]{
\label{fig-alphasf}
\small Results on the flavor independence of
$\alpha_S$ from DELPHI~\cite{bib-delphialphasf}, 
OPAL~\cite{bib-opalalphasf} and SLD~\cite{bib-sldalphasf}.
These results utilize ${\cal{O}}(\alpha_S^2)$ mass
corrections~\cite{bib-masscorr} for c and b quarks.
}
\end{figure}
Previous results on the flavor independence of $\alpha_S$
have been based on leading order mass corrections
for the heavy c and b quarks.
DELPHI~\cite{bib-delphialphasf}, 
OPAL and SLD~\cite{bib-sldalphasf}
have now presented studies
which utilize recent ${\cal{O}}(\alpha_S^2)$ mass
corrections~\cite{bib-masscorr} for c and b quarks.
The results are summarized in Fig.~\ref{fig-alphasf}.
The flavor independence of $\alpha_S$ is verified
to better than 1\% for the b/uds flavors and to 3\%
for the c/uds flavors.

\subsection{Gluon jet studies}

Most studies of gluon jets are from {\epem} annihilations.
Recent results on gluon jets
include a study of unbiased gluon jets by OPAL~\cite{bib-opalgincl},
a study of the scale evolution of gluon jet multiplicity
by DELPHI~\cite{bib-delphigluons},
and a study of the splitting of gluons to 
b$\overline{\mathrm{b}}$ pairs by SLD.
These latter two topics are presented in separate
talks at this conference
(see the contributions by Oliver Klapp and Toshinori Abe,
these proceedings);
hence only the OPAL study is discussed here.

To test theoretical predictions in a meaningful manner,
the experimental definition of jets should match the
theroetical one.
Theoretical predictions of gluon jet properties
are based on the production of a virtual gluon jet
pair, {\gluglu},
from a color singlet point source.
The jet properties are obtained inclusively
by summing over the event hemispheres:
thus there is no selection of a specific event topology.
In contrast,
most studies of gluon jets
employ a jet finding algorithm to identify 
{\epem}$\rightarrow\,${\qq}g
events with a prominent three-jet structure,
interpreted as arising from two quark jets and a gluon jet.
Results from these ``3-jet events'' cannot be used
to test QCD predictions in a quantitative manner
since the experimental selection does not satisfy the
inclusive requirements of the calculations.
In particular,
results based on 3-jet events usually
exhibit a strong dependence on the jet finding algorithm
employed for the analysis.

{\gluglu} production from a color singlet point source
is a process which has been practically unobserved in nature.
One channel where the experimental selection of gluon jets
matches the theoretical criteria is
{\epem} hadronic annihilation events
in which the quark jets q and {\qbar} from the electroweak 
Z$^0/\gamma$ decay are approximately colinear:
the gluon jet hemisphere against which the 
q and {\qbar} recoil is produced under the same
conditions as gluon jets in {\gluglu} 
events~\cite{bib-valery88,bib-gary94}.
OPAL selected events 
of the type {\epem}$\rightarrow\,$q$_{\mathrm{tag}}
\overline{\mathrm{q}}_{\mathrm{tag}}${\gincl},
in which {\gincl} refers to a gluon jet hemisphere
recoiling against two tagged quark jets q$_{\mathrm{tag}}$
and $\overline{\mathrm{q}}_{\mathrm{tag}}$ in the
opposite hemisphere.
The OPAL result is
obtained for a for {\gincl} jet energy of {\ejet}=40~GeV.
\begin{figure}[t]
\vspace*{-1.4cm}
\begin{center}
\begin{tabular}{cc}
\epsfxsize=2.5 truein
\epsffile[55   150 585 750]{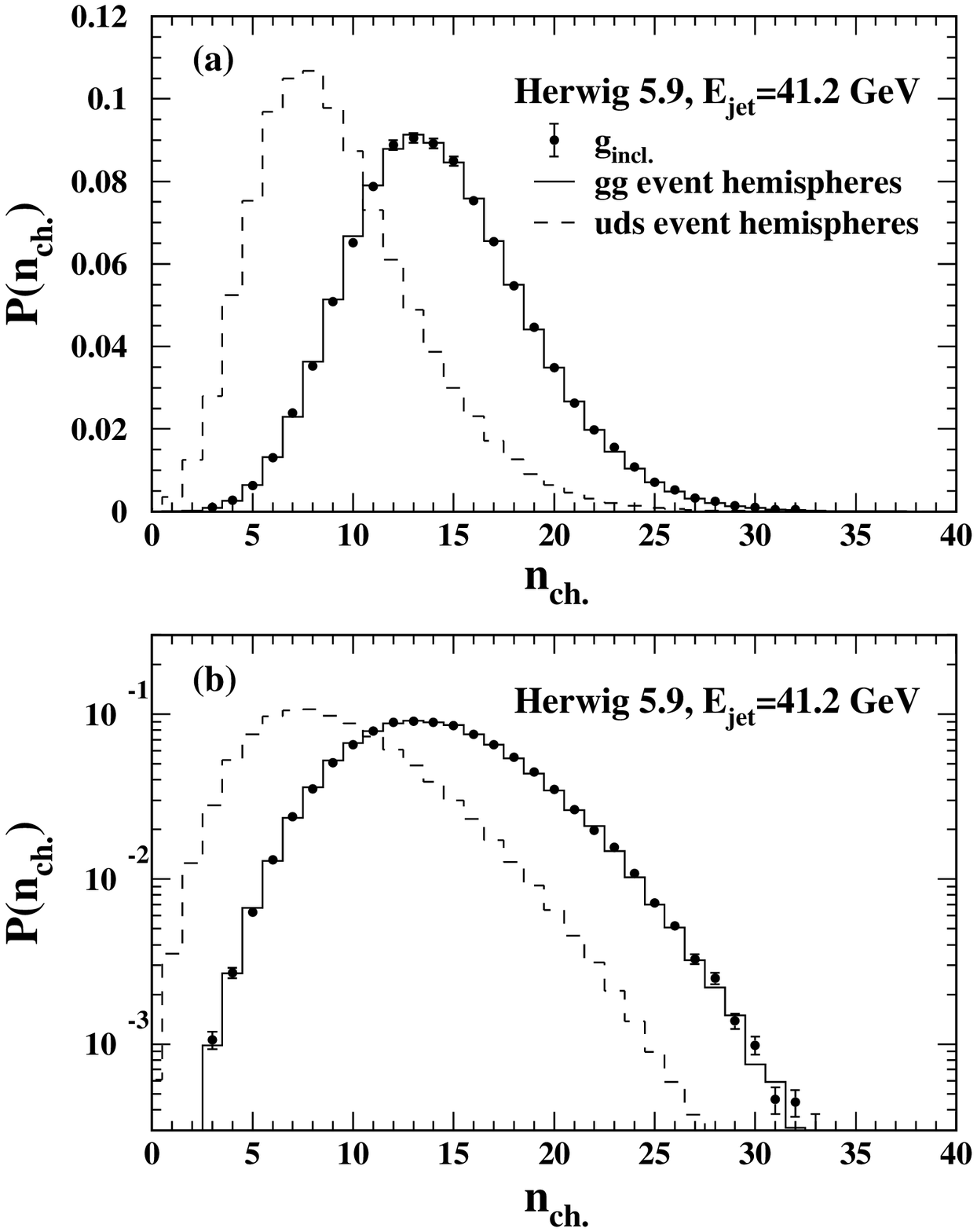} &
\epsfxsize=2.5 truein
\epsffile[55   150 585 750]{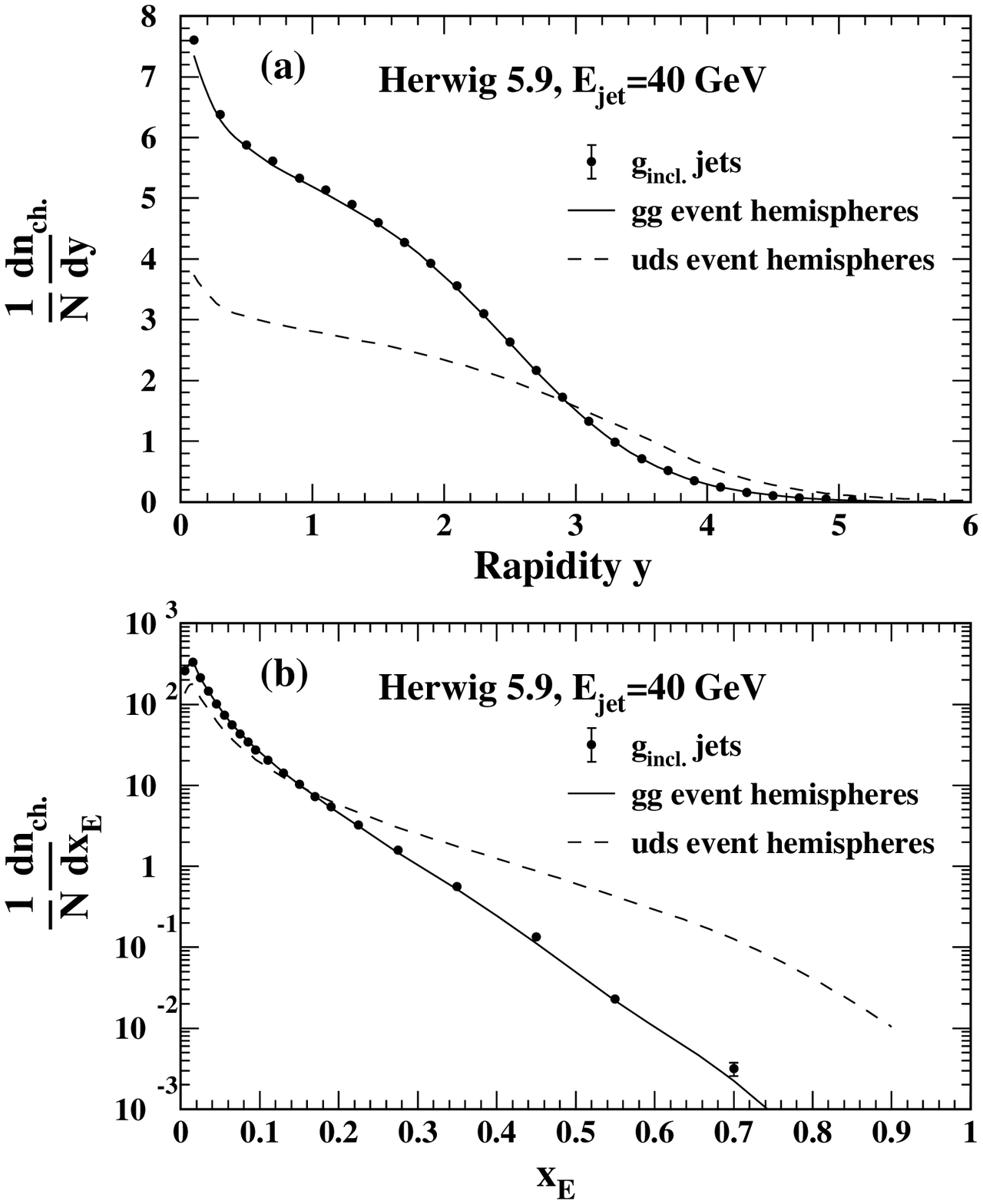} \\
\end{tabular}
\end{center}
\caption[]{
\label{fig-gincl}
\small 
The prediction of the Herwig parton shower
Monte Carlo event generator for {\gincl} gluon jet
hemispheres from {\epem} annihilations,
in comparison to the Herwig predictions for hemispheres
in {\gluglu} and uds events produced from a
color singlet point source~\cite{bib-opalgincl}.
}
\end{figure}
Fig.~\ref{fig-gincl} shows a Monte Carlo
comparison of {\gincl} hemispheres from {\epem} annihilations
and hemispheres of {\gluglu} events,
for the distributions of
charged particle multiplicity~{\nch}, rapidity~$y$,
and scaled particle energy {\xee}$\,$=$\,$$E/E_{\,\mathrm{jet}}$.
The solid points show Monte Carlo predictions for {\gincl} jets.
The solid curves show Monte Carlo predictions
for {\gluglu} event hemispheres with the same energy
as the {\gincl} jets.
The results for {\gincl} jets
and {\gluglu} event hemispheres are 
almost indistinguishable,
establishing the validity of this technique
to identify gluon jets in a manner which corresponds
to point source production from a color singlet.

In the most recent OPAL study of {\gincl} jets~\cite{bib-opalgincl},
439 gluon jets are identified with a purity of~83\%.
The gluon jet hemispheres are compared to hemispheres of
light quark (uds) events,
selected as explained above for the analysis on
the flavor independence of~$\alpha_S$.
The ratio $\rch$ of the mean multiplicity
in gluon jets to that in quark jets is measured to be
$\rch$$\,$=$\,$$1.514\pm 0.019\,(\mathrm{stat.})
\pm 0.034\,(\mathrm{syst.})$,
in excellent agreement with recent QCD calculations
of this quantity~\cite{bib-lupia}.

\begin{figure}[ht]
\vspace*{-1.4cm}
\begin{center}
\begin{tabular}{cc}
\epsfxsize=2.5 truein
\epsffile[55   150 585 750]{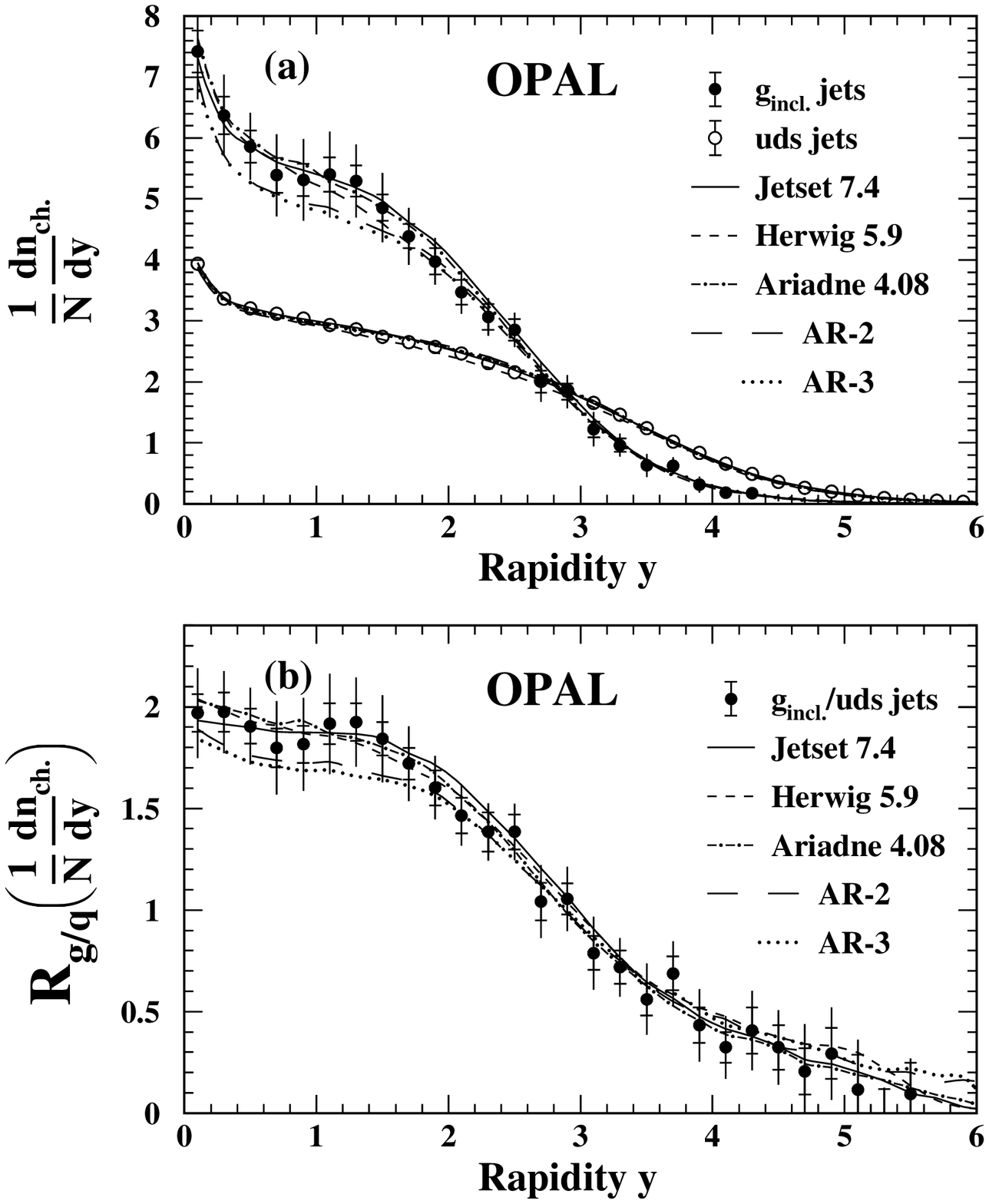} &
\epsfxsize=2.5 truein
\epsffile[55   150 585 750]{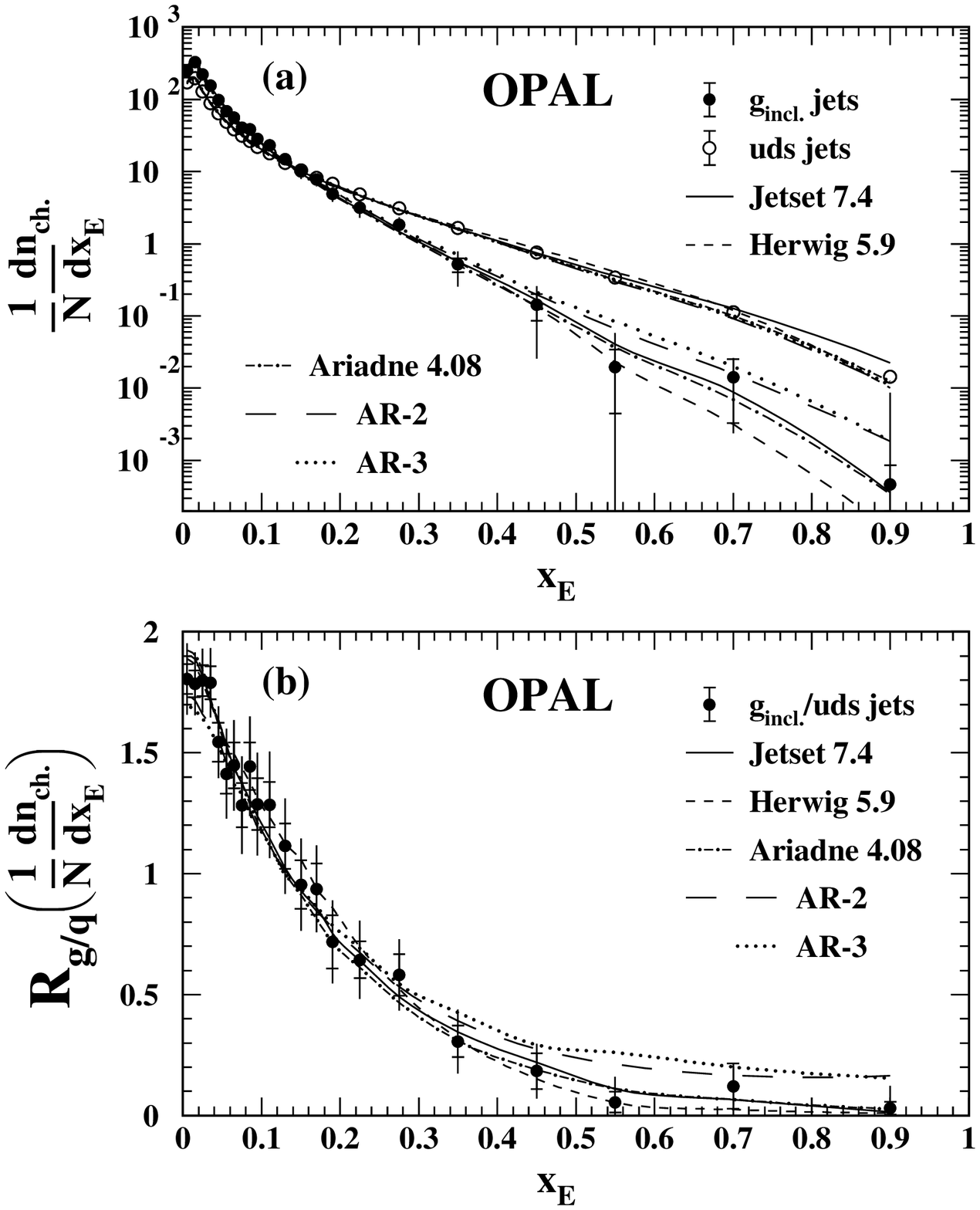} \\
\end{tabular}
\end{center}
\caption[]{
\label{fig-gincl2}
\small 
(Left) Corrected distributions of charged particle rapidity, {\ysph},
for 40.1~GeV {\gincl} gluon jets
and 45.6~GeV uds quark jets.
The ratio of the gluon to quark jet rapidity
distributions for 40.1~GeV jets.
(Right) The corresponding results for charged particle
scaled energy, {\xee}$\,$=$\,$$E/E_{\,\mathrm{jet}}$.
The top plots show the separated gluon and quark jet results;
the bottom plots show the ratio between the two~\cite{bib-opalgincl}.
}
\end{figure}

The measured distributions of
{\ysph} and {\xee} for the gluon and uds hemispheres
are shown in Fig.~\ref{fig-gincl2}.
A striking feature of these results
is the nearly factor of two difference between the
mean multiplicities of gluon and quark jets
at small values of rapidity and~{\xee}.
The ratio of the mean gluon to quark
jet charged particle multiplicity for
{\mysph}$\,\leq\,$1 is measured to be
$\rchysphone$=$1.919\pm 0.047\,(\mathrm{stat.})
\pm 0.095\,(\mathrm{syst.})$.
For {\it soft} particles,
i.e.~particles with energies $E$$\,<<\,${\ejet},
QCD predicts that the mean multiplicities
in gluon and quark jets differ by a factor of 
$r$$\,$=$\,${\cacf}$\,$=$\,$2.25~\cite{bib-stan}.
Monte Carlo study 
demonstrates that the experimental variable
$\rchysphone$ does indeed yield 2.25 at the parton
level for a large energy ({\ejet}=5~TeV):
hence the experimental variable $\rchysphone$
corresponds
to the multiplicity ratio $r$ between gluon and quark jets
as it is defined for analytic calculations.
Because the experimental definition of gluon jets
in this analysis corresponds to the theoretical one,
these results provide the most direct test to date of
QCD predictions for gluon jets.

\vspace*{-.3cm}
\section{Energy scaling of QCD variables}
\label{sec-escaling}

\subsection{Power law corrections for non-perturbative effects}

\begin{figure}[t]
\vspace*{-2.2cm}
\begin{center}
\begin{tabular}{cc}
\epsfxsize=2.5 truein
\epsffile[25   150 555 750]{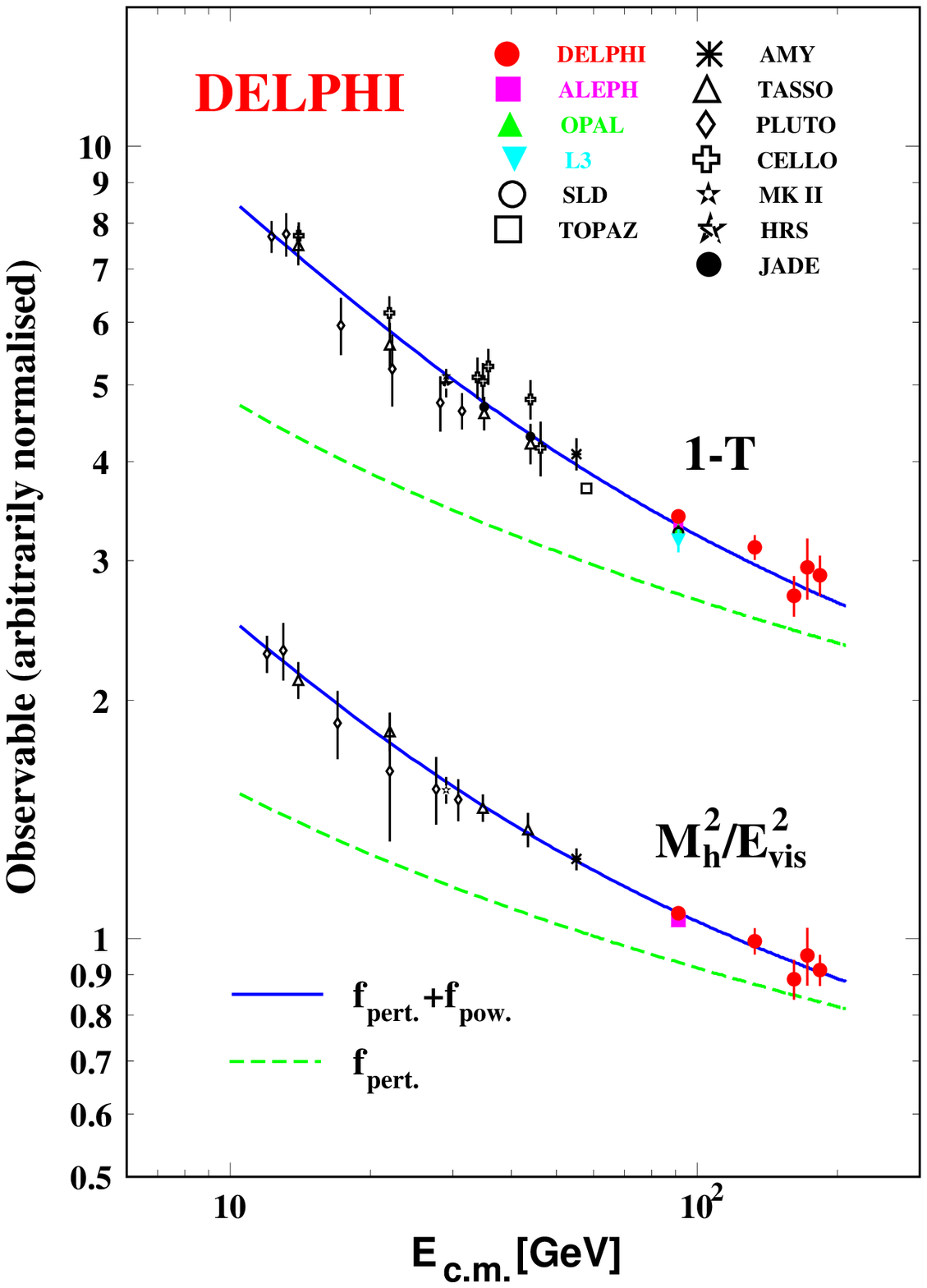} &
\epsfxsize=2.8 truein
\epsffile[25   230 555 830]{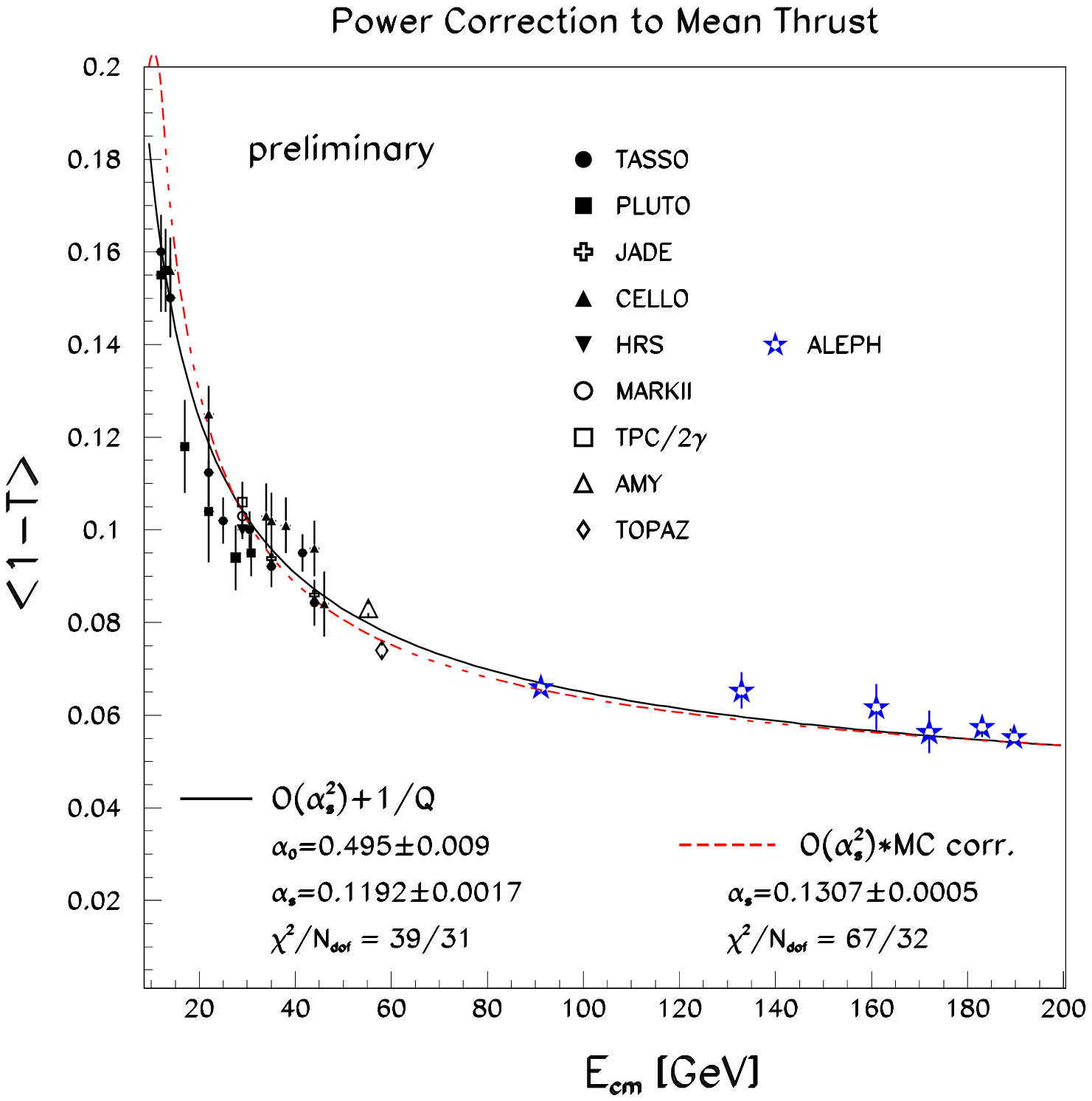}
 \\
\end{tabular}
\end{center}
\vskip -.5 cm
\caption[]{
\label{fig-power}
\small Results on the energy evolution of the mean values of
Thrust and Heavy Jet Mass in {\epem}
annihilations~\cite{bib-power}.
}
The preliminary results for $\alpha_S$ and $\alpha_0$
from DELPHI and ALEPH are given in the table below.
The fits yield generally
consistent results for $\alpha_S$(M$_{\mathrm Z}$)
and $\alpha_0$:
in this sense the power law ansatz provides a
successful description of the data.
\end{figure}

A power law ansatz has been presented~\cite{bib-dokshitzer}
to treat hadronization corrections analytically.
The ansatz takes the form
\begin{equation}
  \langle y \rangle = \langle y_{pert.} \rangle + \langle y_{non-pert.} \rangle
\label{eq-power}
\end{equation}
where $\langle y\rangle$ represents the mean value
of an event shape variable such as Thrust,
with $\langle y_{pert.} \rangle$ a term calculable in
perturbative QCD,
and $\langle y_{non-pert.} \rangle$ a non-perturbative term
meant to replace the Monte Carlo derived hadronization
corrections often used in the experimental analysis
of {\epem} data.
The perturbative term $\langle y_{pert.} \rangle$
has been calculated to ${\cal{O}}(\alpha_S^2)$.
The non-perturbative term is given by
$$
   \langle y_{non-pert.} \rangle 
   = \frac{C_yf(\alpha_S,\alpha_0)}{E_{\,\mathrm{c.m.}}}
$$
where $f$ is a universal function of $\alpha_S$ and $\alpha_0$,
with $\alpha_0$ a non-perturbative parameter
predicted to have the same value for all variables~$y$.
This ansatz has been tested by DELPHI and ALEPH
by fitting expression~(\ref{eq-power}) to {\epem} measurements
of $\langle y\rangle$ versus {\ecm},
with $\alpha_S$ and $\alpha_0$
as fitted parameters.
The renormalization scale is chosen to be the mass of the Z$^0$
for this fit.
The DELPHI study,
using LEP and lower energy {\epem} data,
is summarized in Fig.~\ref{fig-power} (left).
DELPHI base their analysis on the energy evolution of the mean
Thrust and Heavy Jet Mass values.
The solid lines in Fig.~\ref{fig-power} (left)
show the results of the fit.
The contribution of the perturbative term is shown by the
dashed lines.
The corresponding ALEPH results for the mean Thrust value
are shown in Fig.~\ref{fig-power} (right).
ALEPH also includes the mean
C-parameter~\cite{bib-cparameter} value in their study.

\vspace*{-.9cm}
\subsection{Energy scaling of {\boldmath $\alpha_S$} and mean multiplicity}

\begin{figure}[ht]	
\vspace*{-1cm}
\begin{center}
\begin{tabular}{cc}
\epsfxsize=2.5 truein
\epsffile[55   150 585 750]{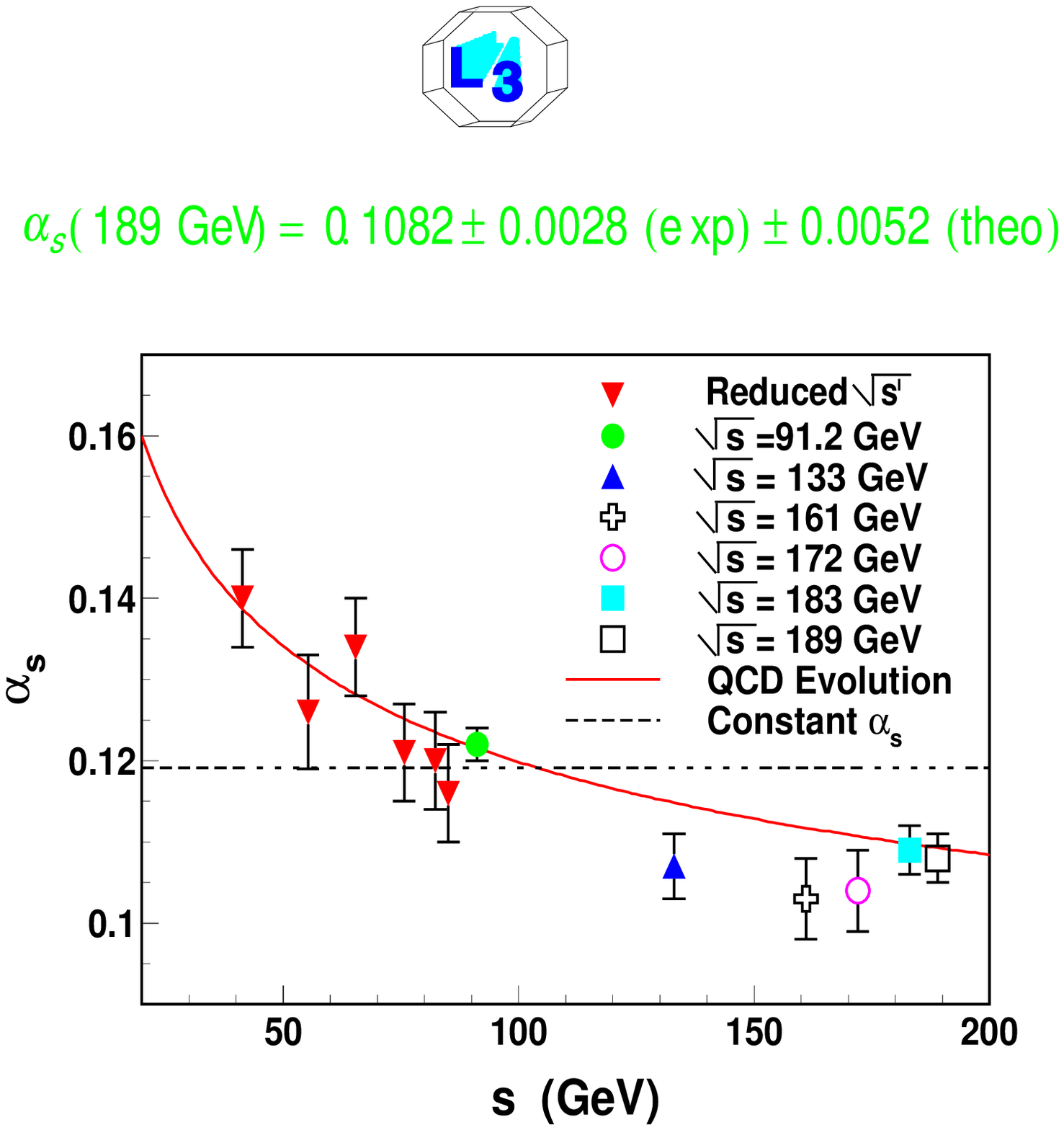} &
\epsfxsize=2.5 truein
\epsffile[55   150 585 750]{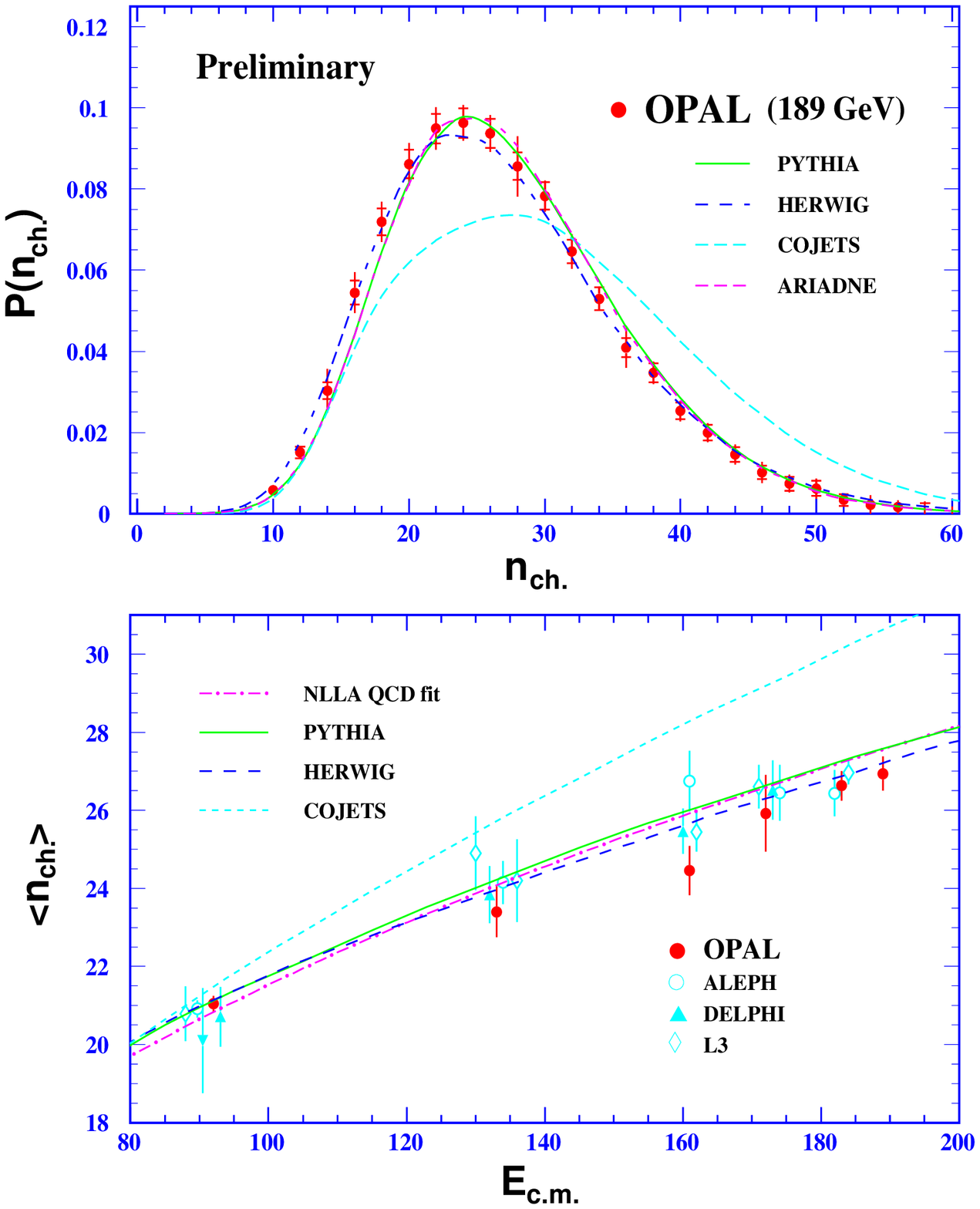} \\
\end{tabular}
\end{center}
\vspace*{-.3cm}
\caption[]{
\label{fig-escaling}
\small 
(Left) L3 results on the running of $\alpha_S$~\cite{bib-clare98}.
(Right) OPAL measurements the charged particle multiplicity
distribution at 189~GeV and of the energy evolution of the
mean charged multiplicity~\cite{bib-plane98}.
}
\end{figure}

Last,
we present studies of the energy scaling of $\alpha_S$ and
{\mnch} at LEP.
Fig.~\ref{fig-escaling} (left) shows measurements of $\alpha_S$
from L3.
The measurements for {\ecm}$>$M$_{\mathrm{Z}}$
are determined using the LEP-1, LEP-1.5 and LEP-2 data
in the manner described in section~\ref{sec-189}.
The measurements below {\ecm}=M$_{\mathrm{Z}}$
are obtained from LEP-1 events
in which the initial-state e$^-$ or e$^+$
radiates a photon,
thus reducing the energy of the hadronic system.
The uncertainties shown are statistical only,
i.e.~the correlated systematic terms are omitted to
emphasize the running character of~$\alpha_S$.
The solid line shows the evolution predicted by QCD,
which agrees well with the data.
It is interesting that the high luminosity 183 and 189~GeV
points fall directly on the solid curve which also passes
through the high statistics LEP-1 point.
Fig.~\ref{fig-escaling} (right) shows measurements of {\mnch}
versus~{\ecm}.
The top portion of this figure shows the {\nch} distribution measured 
by
\begin{table}[tb]
\centering
\begin{tabular}{ccccc}
  & & & & \\[-2.4mm]
 Experiment & Event Shape Variable
  & $\alpha_0$ & $\alpha_S$(M$_{\mathrm Z}$) 
  & $\chi^2$/d.o.f. \\[2mm]
\hline
\hline
DELPHI & T & $0.49\pm0.1$ & $0.119\pm0.005$ & 1.9 \\
DELPHI & (M$_{\mathrm{H}}$/E$_{\,\mathrm{c.m.}})^2$ &
     $0.55\pm0.03$ &  $0.119\pm0.004$ & 0.2 \\
ALEPH & T & $0.45\pm0.08$ & $0.119\pm0.006$ & 1.3 \\
ALEPH & C & $0.46\pm0.06$ & $0.113\pm0.004$ & 0.8 \\ 
\end{tabular}
\label{tab-power}
\end{table}

\noindent
OPAL at 189~GeV.
Shown in comparison to the data are the predictions of the
principle QCD Monte Carlo programs,
tuned to provide an approximately equivalent description of global
event properties at the Z$^0$.
Pythia, Herwig and Ariadne,
all based on parton showers with coherence
(soft gluon interference),
are seen to describe the energy
evolution of the mean multiplicity well.
Cojets~\cite{bib-cojets},
based on a parton shower without coherence,
describes the high energy data poorly and thus does not
exhibit the correct energy scaling behavior.
A similar result is obtained by ALEPH using a variant
of Pythia without coherence~\cite{bib-lancon98}.
These results are highly suggestive of the need to
include coherence effects in QCD predictions to obtain
an accurate description of multiplicity in {\epem} data.

\end{document}